\documentclass[nofootinbib,twocolumn,showpacs,amsmath,amssymb]{revtex4}
\usepackage{amsmath}
\usepackage{amssymb}
\usepackage{amsfonts}
\usepackage{epsfig}
\begin{document}

\title{Space-time non-commutativity tends to create bound states}

\author{Dmitri V. Vassilevich}
\email{Dmitri.Vassilevich@itp.uni-leipzig.de}
\affiliation{Institut f\"{u}r Theoretische Physik, Universit\"{a}t
Leipzig, Augustusplatz 10, D-04109 Leipzig, Germany}
\affiliation{V.~A.~Fock Institute of Physics, St.Petersburg University,
Russia}
\author{Artyom Yurov}\email{yurov@freemail.ru, artyom_yurov@mail.ru}
\affiliation{Department of Theoretical Physics,
Kaliningrad State University,
236041, A. Nevskogo 14,
Kaliningrad,
Russia}

\begin{abstract}
We study the spectrum of fluctuations about static solutions
in $1+1$ dimensional non-commutative
scalar field models. In the case of soliton
solutions non-commutativity leads to creation of new bound states.
In the case of static singular solutions an infinite tower of
bound states is produced whose spectrum has a striking similarity
to the spectrum of confined quark states.
\end{abstract}

\maketitle

\section{Introduction}
Over the recent years non-commutative field theory has developed
into a mature discipline (see reviews \cite{reviews}).
It was argued (cf., e.g., \cite{non-unitar}; more references
can be found in
\cite{SYL})
that because of the presence of an infinite
number of time derivatives space-time non-commutative theories 
cannot be quantized properly. However, the situation does not look
hopeless. Perturbative unitarity can be successfully maintained 
\cite{SYL} if one takes care of explicit Hermiticity of the
Lagrangian. Even a canonical formalism can be developed at the expense
of introducing an additional space-time dimension 
\cite{Gomis:2000gy}. We don't have much to add to this discussion.
Moreover, our analysis will be essentially classical. We like to mention
only that space-time non-commutative theories are not excluded, and that
one can expect many non-standard features from these theories. 

In this paper we consider some qualitative features of space-time
non-commutative theories. Namely, we study fluctuations around
static classical solutions in $1+1$-dimensional
non-commutative models with a  real scalar field. 
Note, that the solutions themselves
look exactly as in the commutative models. Therefore, non-commutativity
can be seen only through the fluctuation spectra or through the scattering
amplitudes \cite{Valtancoli:2003vy}. We find that the frequency-dependent
potential which appears in the equation for fluctuation has typically
an ``effective width'' proportional to the frequency and to the 
non-commutativity parameter. This phenomenon is somewhat similar to
delocalization of states discussed in \cite{Dubovsky:2002bv} in a different
context. In our case, this distortion of the potential leads to creation
of new bound states (soliton backgrounds) or even to infinite families
of new bound states (singular static backgrounds).

This paper is organized as follows. In the next section we fix our
notations and conventions. Section \ref{solsec} is devoted to fluctuations
about solitonic solutions in the $\phi^4_\star$ and in  sine-Gordon models.
Singular solutions are discussed in section \ref{singsec}. Some concluding
remarks are given in section \ref{consec}.
\section{Notations and conventions}
Let us consider the non-commutative plane with a coordinate  $\sigma =(t,x)$.
The Groenewold-Moyal product is defined by the equation
\begin{equation}
 f(\sigma )\star g(\sigma)
=\left[\exp\left(\frac{i}{2}
\theta^{mn}\frac{\partial}{\partial \sigma^m}\frac{\partial}
{\partial \sigma^{'n}}\right)f(\sigma)
g(\sigma')\right]_{\sigma'=\sigma},\label{Moyal}
\end{equation}
where $\theta^{mn}=-\theta^{nm}$ is a constant antisymmetric 
$2\times 2$ matrix which can be chosen as $\theta^{mn}=2\theta\epsilon^{mn}$
with
$\epsilon^{mn}=-\epsilon^{nm}$ and  $\epsilon^{01}=1$.
This product is associative but non-commutative. 
A historical overview can be found in \cite{Zachos:2001ux}.

The following relations will be useful throughout this paper:
\begin{equation}
f(x)\star e^{i\omega t}=e^{i\omega t} f(x+\theta\omega),
\qquad e^{i\omega t}\star f(x)=e^{i\omega t} f(x-\theta\omega).
\label{useful}
\end{equation}

We shall study non-commutative deformations of the action
\begin{equation}
S=\int d^2\sigma \left[ 
\frac {1}{2}\partial_{n}\phi\,\partial^{n}\phi-V(\phi) 
\right] \label{action}
\end{equation}
for a real one-component field $\phi$ with some potential $V$.
Non-commutative deformations of $V$ are constructed (as usual)
in the following way. Let 
\begin{equation}
V(\phi )=\sum_{p \ge 0} c_p \phi^p \,.\label{Taylor}
\end{equation}
Then a non-commutative counterpart of $V$ is defined as
\begin{equation}
V_\star (\phi )
=\sum_{p \ge 0} c_p \phi\star\phi \dots \star \phi \,,\label{ncTaylor}
\end{equation}
where the $p$th term contains $p$th star-power of $\phi$. We restrict 
ourselves to polynomial or exponential potential only, so that there is
no problem with the convergence of (\ref{Taylor}). Convergence of
(\ref{ncTaylor}) is a more subtle question, but we shall actually 
work  with non-commutativity only in the perturbative regime.

Our primary example will be the $\phi^4$ model
\begin{equation}
V^{[4]}=-\frac 12 m^2 \phi^2 + \frac {\lambda}4 \phi^4 \,.\label{p4pot}
\end{equation}
We shall also consider the Liouville model\footnote{To avoid
confusions we have to mention that sometimes the Liouville model
includes also an interaction with two-dimensional metric.}
\begin{equation}
V^{[L]}=\gamma e^{\beta\phi}\,,\label{Lipot}
\end{equation}
the sine-Gordon model
\begin{equation}
V^{[sG]}
=-\frac{m^4}{6\lambda} \cos \left( \frac{\sqrt{6\lambda}}{m}\phi\right)\,,
\label{sGpot}
\end{equation}
and the sinh-Gordon model
\begin{equation}
V^{[shG]}= \frac{m^2}2 \cosh (2\phi ) \,.\label{shGpot}
\end{equation}

The equation of motion following from the non-commutative deformation
\begin{equation}
S_\star =\int d^2\sigma \left[ 
\frac {1}{2}\partial_{n}\phi\,\partial^{n}\phi-V_\star (\phi) 
\right] \label{ncaction}
\end{equation}
of (\ref{action}) reads:
\begin{equation}
\partial_t^2 \phi - \partial_x^2 \phi + [\partial_\phi V ]_\star =0\,.
\label{nceom}
\end{equation}
Obviously, star-product of functions depending on $x$ only coincides
with the ordinary product. Therefore, all \textit{static}
solutions of a commutative model will also solve the non-commutative
equation of motion (\ref{nceom}).
Note, that there could be of course non-static localized
solutions in space-time non-commutative theories (cf.
\cite{Bak:2003ua}).
\section{Solitons: new bound states\label{solsec}}
\subsection{The $\phi^4_\star$ model}
In this section we consider the spectrum of fluctuations about 
static solitonic solutions (i.e. 
about localized solutions with finite energy). For the $\phi^4$
model (\ref{p4pot}) this is the kink solution: 
\begin{equation}
\Phi(x)=\frac{m}{\sqrt{\lambda}}\tanh\left(\frac{mx}{\sqrt{2}}\right)
\label{kink}
\end{equation}
As we have already mentioned above, this is also a soliton in the
non-commutative $\phi^4_\star$.

Let us consider small fluctuations about the kink background,
$\phi :=\Phi +\delta \phi$. The equation of motion for the
fluctuations reads
\begin{equation}
\delta\ddot\phi-\delta\phi''
-m^2\delta\phi+\lambda(\delta\phi\star\Phi^2+\Phi^2\star\delta\phi+
\Phi\star\delta\phi\star\Phi)=0.\label{fleq}
\end{equation}
We shall look for the solutions in the form
\begin{equation}
\delta\phi = e^{i\omega t}\eta(x) \,.\label{ans}
\end{equation}
Then, by virtue of (\ref{useful}), one obtains
\begin{equation}
-\eta''+\lambda\left(\Phi_+^2+\Phi_+\Phi_-+\Phi_-^2\right)\eta
=\left(\omega^2+m^2\right)\eta\,,\label{eqeta}
\end{equation}
where $\Phi_{\pm}\equiv\Phi(x_{\pm})=\Phi(x\pm \theta\omega)$.

This problem has a natural scale $\mu = m/\sqrt{2}$. It is convenient
to introduce rescaled dimensionless variables:
$\hat x=\mu x$, $\hat \omega = \omega /\mu$, $\hat\theta =\mu^2 \theta$.
In terms of these variables eq. (\ref{eqeta}) reads
\begin{equation}
-\eta''(\hat x)+U(\hat x)\eta(\hat x)=\left({\hat\omega}^2+2\right)\eta\,
\label{neqeta}
\end{equation}
where prime denotes the differentiation with respect to $\hat x$, and
\begin{equation}
U=2\left(\tanh^2({\hat x}_+)+\tanh({\hat x}_+)\tanh({\hat x}_-)
+\tanh^2({\hat x}_-)\right)\,.
\label{defU}
\end{equation}
From now on we drop hats which is equivalent to setting
$\mu =1$.

In the commutative case, $\theta=0$, $x_\pm =x$,
\begin{equation}
U=U_0=6 -\frac{6}{\cosh^2{ x}} \,.\label{Ucom}
\end{equation}
There are two bound states for this potential with 
$\omega_1=0$ and $\omega_2=\sqrt{3}$ with the wave functions:
\begin{equation}
\eta_1=\frac{\sqrt{3}}{2\cosh^2{x}},\qquad 
\eta_2=\sqrt{\frac{3}{2}}\frac{\sinh{ x}}{\cosh^2{ x}}
\label{combound}
\end{equation}

\begin{figure}[ht]
\epsfxsize=8cm\epsffile{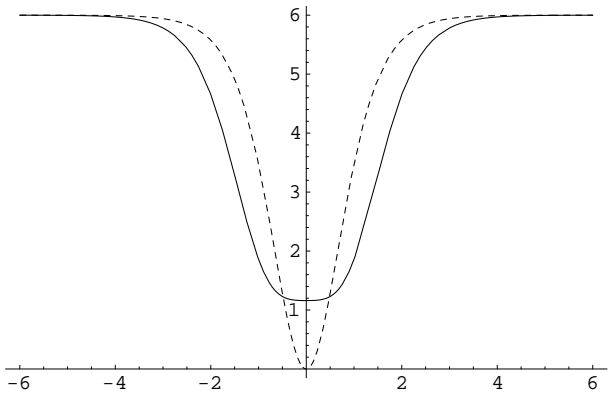}
\caption{The potential $U(x)$ for (dimensionless) $\theta\omega=1$
 (solid line) as compared to $U_0$ (dashed line).}
 \label{smtheta}
\end{figure}

For $\theta \ne 0$ we have a complicated problem with a frequency
dependent potential. Of course, in the generic case no exact solution
for the eigenstates is available. For small $\theta$ and relatively low
eigenfrequencies, $\theta \omega \ll 1$, we can use ordinary perturbation
theory to estimate shift of the eigenvalues. 
An example of the potential $U(x)$ is shown on Fig.\ \ref{smtheta}.

We write to the leading
order in $\theta\omega$:
\begin{equation}
-\eta''-\left(\frac{6}{\cosh^2x}-U^{(1)}\right)\eta
=\left(\omega^2-4\right)\eta,
\label{pert}
\end{equation}
where
\begin{equation}
U^{(1)}=\frac{2(\theta\omega)^2}{\cosh^2x}\left(\frac{7}{\cosh^2x}-6\right).
\label{U1}
\end{equation}
We assume $\eta =\eta_0 +\delta \eta$, where $\eta_0$ satisfies
(\ref{neqeta}) with $U=U_0$ given by (\ref{Ucom}). Remember that hats
have been dropped. Then one immediately obtains that the bound state
frequency $\omega_1=0$ is not shifted to this order, while $\omega_2$
receives
a negative correction:
\begin{equation}
{\omega}_2^2=\omega_2^2\vert_{\theta=0}+
\int dx \eta_2^2 U^{(1)}=3\left(1-\frac{8}{5}\theta^2\right)
\label{om2}
\end{equation}
or
\begin{equation}
\omega_2^2=(1-0.8\theta^2)\omega_2\vert_{\theta=0}\,.
\label{om22}
\end{equation}
Corrections to this formula are of order $\theta^4$.
We see, that in this regime the eigenfrequency of one of the
bound states is being shifted. Let us remind that the kink solution
itself does not depend on $\theta$. Therefore, shift of the eigenfrequencies
may be a measurable manifestation of non-commutativity for small
$\theta$ and low frequencies.   

We have shown that $\omega$ decreases due to the non-commutativity.
A natural question to ask is whether \textit{new} bound states
can appear. The answer is positive. The following analysis will
be made in the large-$\theta$ limit.

\begin{figure}[ht]
\epsfxsize=8cm\epsffile{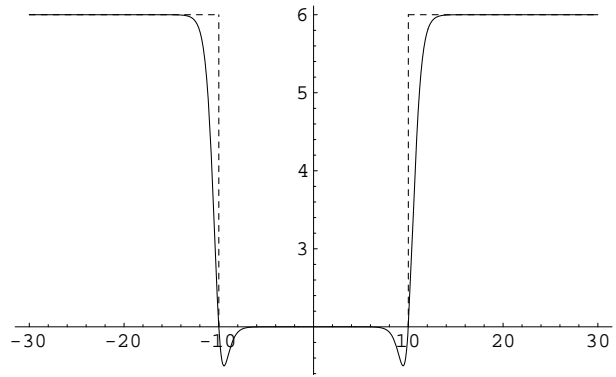}
\caption{The potential $U(x)$ 
(solid line) and the approximating square well potential $\tilde U$ 
(dashed line) for $\theta\omega=10$.}
 \label{ltheta}
\end{figure}

It is easy to demonstrate that for large $\theta\omega$ the potential
$U(x)$ (cf. (\ref{defU})) can be approximated by a square well potential
$\tilde U$ (cf. Fig.\ \ref{ltheta}) such that 
\begin{eqnarray}
&&\tilde U(x)=6 \quad \text{for} \quad |x| > \theta\omega \,,\nonumber\\
&&\tilde U(x)=2 \quad \text{for} \quad |x| < \theta\omega \,.\label{tilU}
\end{eqnarray} 
Therefore, we replace (\ref{neqeta}) by
\begin{equation}
-\eta''+\tilde U \eta =\left( \omega^2 +2 \right) \eta \,.
\label{tileq}
\end{equation}
Note, that the characteristic width of $\tilde U$ depends on $\omega$.
Clearly, bound states can only appear for $0<\omega <2$. By using the standard
methods which can be found in any quantum mechanics text book we obtain
that eigenfrequencies of the bound states should satisfy one of the
equations:
\begin{equation}
\tan(\theta\omega^2)=\frac{\sqrt{4-\omega^2}}{\omega},\qquad 
\cot(\theta\omega^2)=-\frac{\sqrt{4-\omega^2}}{\omega}\,,\label{twoeqs}
\end{equation}
where the first equation gives eigenfrequencies of the states with
a symmetric wave function, while the second equation describes
the states with antisymmetric wave functions.

In our approximation it is essential that $\omega\theta$ is large.
Therefore, we shall consider (\ref{twoeqs}) for $\omega$ near the
upper limit $\omega =2$. In a small interval near $\omega^2=4$
the functions on the right hand sides of the equations (\ref{twoeqs})
are bounded and continuous, while $\tan(\theta\omega^2)$ and
$\cot(\theta\omega^2)$ change from $-\infty$ to $+\infty$
when $\omega^2$ changes from 
$\pi (n - \frac 12) /\theta$ to $\pi (n + \frac 12) /\theta$
or from $\pi n /\theta$ to $\pi (n - 1) /\theta$
respectively. This means if that $\theta \gg \pi/8$ there is always
at least one solution for each of the equations (\ref{twoeqs})
near $\omega =2$. We can even estimate roughly the number of the
solutions in the upper half of the allowed interval (i.e. for
$\omega^2 \in [2,4]$) to be about $8\theta /\pi$. 

We conclude that for a large non-commutativity parameter $\theta$
there are many new bound states for the fluctuations about the
kink soliton as compared to the commutative case.  

\subsection{The sine-Gordon model}
To make sure that the phenomenon of
creation of new bound states due to the non-commutativity
is present not only in the $\phi^4_\star$ model, let us consider
the sine-Gordon model (\ref{sGpot}). Static solutions in this
model in both commutative and non-commutative regimes should
satisfy the equation:
\begin{equation}
-\phi''+\frac{m^3}{\sqrt{6\lambda}}
\sin\left(\frac{\sqrt{6\lambda}}{m}\phi\right)=0.
\label{sGeq}
\end{equation}
There is a one-soliton solution of (\ref{sGeq}) which reads
\begin{equation}
\Phi(x)=\frac{4m}{\sqrt{6\lambda}}\arctan\left(e^{mx}\right)\,.
\label{sGsol}
\end{equation}

\begin{figure}[ht]
\epsfxsize=8cm\epsffile{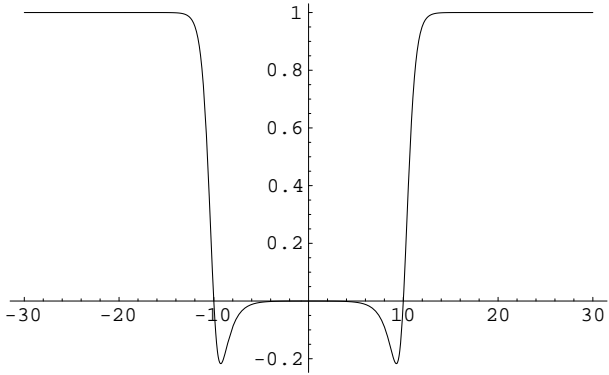}
\caption{The effective potential (\ref{sGflu}) for the sine-Gordon model
 in dimensionless 
variables ${\hat x}=mx$, ${\hat\theta}=m^2\theta$, ${\hat\omega}=\omega/m$ at 
${\hat\theta}{\hat\omega}=10$.}
\label{sGplot}
\end{figure}

To obtain an equation for fluctuations we have to expand a non-commutative
exponential. This can be done with the help of the equation:
\begin{equation}
e^{A+B}_\star = e^A_\star + \int_0^1 d\sigma\,e^{\sigma A}_\star
\star B\star  e^{(1-\sigma)A}_\star + \mathcal{O} (B^2)
\,,\label{exex}
\end{equation}
which has a purely combinatorial origin and is true regardless of the
choice of associative product involved (this could be the ordinary
operator or matrix product, for example, or the Groenewold-Moyal star
as in our case). The formulae (\ref{exex}),
(\ref{ans}) and (\ref{useful}) yield
\begin{equation}
-\eta''+\frac{m^3\left(\sin\left(\frac{\sqrt{6\lambda}}
{m}\Phi_+\right)-\sin\left(\frac{\sqrt{6\lambda}}{m}\Phi_-\right)\right)}
{\sqrt{6\lambda}(\Phi_+-\Phi_-)}\eta=\omega^2\eta\,.
\label{sGflu}
\end{equation}
For large $\theta\omega$ the effective potential $U(x)$ behaves
similarly to that for the $\phi^4_\star$ model (see Fig.\ \ref{sGplot}).
Namely, the potential $U(x)$ can be approximated by a square well
potential with the width $2\hat\theta \hat \omega$.  
All arguments of the previous subsection apply for this case
almost without modification. We conclude, that for large
non-commutativity we have new bound states. This seems to be
a generic feature of the fluctuation equation on the background
of a static solitonic solution in a two-dimensional non-commutative
space-time.

\section{Singular solutions and confining potentials\label{singsec}}
\subsection{Massless $\phi^4_\star$ model}
\begin{figure}[ht]
\epsfxsize=8cm\epsffile{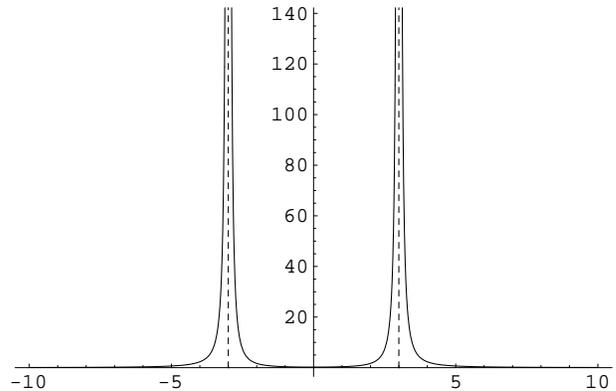}
\caption{The potential $U$ from eq.\ (\ref{p4sflu}) for $\theta\omega\sim 3$.}
 \label{singplot}
\end{figure}

The $\phi^4_\star$ model with $m=0$ admits a singular
static solution which reads:
\begin{equation}
\Phi(x)=\frac{\sqrt{2}}{x\sqrt{\lambda}}. \label{p4ss}
\end{equation}
Then, by acting exactly as in the previous section we obtain
the following equation for the fluctuations:
\begin{equation}
-\eta''+\frac{2(3x^2+\theta^2\omega^2)}{(x^2-\theta^2\omega^2)^2}\eta
=\omega^2\eta.
\label{p4sflu}
\end{equation}
Note, that this equation is scale-invariant, i.e., if we re-scale
$x\to x\mu$, $\omega \to \omega /\mu$, $\theta \to \theta\mu^2$
the scaling parameter cancels out. As a consequence, we can assume that
we are working in dimensionless variable, so that $\theta\omega =3$
on Fig.~\ref{singplot} indeed makes sense.

Obviously, for $\theta =0$ there are no bound states. If $\theta\ne 0$
the situation changes drastically (cf. Fig.\ \ref{singplot}).
In this case we have two infinitely high potential barriers
located at $x=\pm \omega\theta$. Physics between these two barriers
can be approximated by an infinitely deep well of width $2\theta\omega$.
This approximation is good for high frequencies. In this case
the equation for small fluctuations can be easily solved yielding 
\begin{equation}
\omega_{_N}\simeq \sqrt{\frac{\pi N}{2\theta}}, \label{p4spec}
\end{equation}
where $N\in \mathbb{N}$. Accuracy of this formula increases for
large $N$ and/or large $\theta$.

The spectrum obtained in this simple model has a striking similarity
to the spectrum of hadrons. Indeed, we observe an infinite number of
bound states with a linear dependence
of $\omega^2 :=M^2$ on an integer spectral parameter. Since we do not have
something like angular momentum in two dimensions we cannot push
these arguments further. 
\subsection{Exponential interactions}
Again, we would like to test the observation made for the $\phi^4_\star$
model by considering other models admitting similar types of the
classical solution. We start with the Liouville model (\ref{Lipot}).
Classical equation of motion for this model reads
\begin{equation}
\ddot\phi-\phi''+\alpha e^{\beta\phi}_\star =0\,,
\label{Leq}
\end{equation}
where $\alpha = \beta\gamma$. In the commutative case $\theta =0$
there is a general solution to this equation:
\begin{equation}
\phi=\frac{1}{\beta}\log\left(-\frac{2G'(p)F'(q)}{
\alpha\beta\left(G(p)+F(q)\right)^2}\right),
\label{Lsol}
\end{equation}
with $p=(t+x)/2$, $q=(t-x)/2$. $G$ and $F$ are arbitrary functions.
The simplest static solution (which
again is common for the commutative and non-commutative models) is
obtained by setting $G=p$, $F=-q$:
\begin{equation}
\Phi(x)=-\frac{1}{\beta}\log\left(\frac{\alpha\beta x^2}{2}\right), 
\qquad \alpha\beta>0.
\label{sLsol}
\end{equation}

Next we use again the expansion (\ref{exex}), the ansatz
(\ref{ans}), and the property (\ref{useful}) to write down
the equation for fluctuations:
\begin{equation}
-\eta''+\alpha\left(\frac{e^{\beta \Phi_+}
-e^{\beta \Phi_-}}{\Phi_+-\Phi_-}\right)\eta=\omega^2\eta.
\label{Lflu}
\end{equation}
The substitution of (\ref{sLsol}) in (\ref{Lflu}) yields the following
effective potential
\begin{equation}
U=\frac{8\theta\omega x}{(x_+x_-)^2\log\left(\frac{x_+}{x_-}\right)^2}.
\label{LU}
\end{equation}
This potential is similar to the one appearing in the $\phi_\star^4$ model
(cf. (\ref{p4sflu}). Again, we have two infinitely hight potential
barriers with a ``confinement'' region between them. Effective width of
this region is $2\theta\omega$. Therefore, the spectrum of higher excited
states is again given by (\ref{p4spec}).
 
A very similar behaviour can be found also in the sinh-Gordon model
(\ref{shGpot}) near the static solution
\begin{equation}
\Phi(x)=\frac{1}{2}\log\tanh^2\left(\frac{mx}{2}\right).
\label{shGsol}
\end{equation}
We leave this case for the reader as an exercise.

We like to stress that in each case the universal formula (\ref{p4spec})
appears.
\section{Conclusions\label{consec}}
Our main result is that in the presence of the space-time non-commutativity
the effective potential describing fluctuations on a static background
becomes delocalized with the effective width $\sim \theta\omega$.
As a consequence, in the case of large non-commutativity $\theta$
we have much more bound states on solitonic background then in
corresponding commutative theories. On singular static backgrounds
the picture is even more interesting. Non-commutativity produces an
infinite tower of bound states with linear dependence of $\omega^2$
on an integer quantum number (for large frequencies). This behaviour
is universal, the large frequency spectrum depends on $\theta$ only,
but not on the details of the models. This result suggests that the
space-time non-commutativity may have some relation to the problem
of quark confinement. Although we have not presented any general
proof, the number of examples considered seems to justify our
conclusion that creation of new bound states is a generic feature
of space-time non-commutative theories in $1+1$ dimensions.

In principle, the frequency spectrum can be used to calculate quantum
corrections to the mass of the solitons in non-commutative theories.
In \cite{Vassilevich:2003yz} it was argued that the $\zeta$-function
or other heat kernel based methods may be a suitable instrument
(although, it is not clear whether the fluctuation operator involved
can indeed be reduces to $\star$-Laplacians considered in 
\cite{Vassilevich:2003yz}). As an intermediate step one has to put the
system in a large box so that the spectrum becomes discrete. At the end of
the calculation the 
boundary is moved to the infinity and, if necessary, the boundary contribution
to the vacuum energy is subtracted from the total energy of the system
(cf. \cite{Bordag:2002dg} where this procedure is applied to the kink).
However, in the present case no fixed boundary can be far away enough
since the ``effective width'' of the potential is proportional to
the frequency. This seems to be another manifestation of the mixing
between ultra violet and infra red scales in non-commutative theories
\cite{UVIR}. Consequences of this mixing for quantum corrections
to (space) non-commutative solitons in $2+1$ dimensions were considered
recently in  \cite{Xiong:2002tk}. 
\begin{acknowledgments}
This work was supported in part by the DFG Project BO 1112/12-1,
by the Erwin Schr\"{o}dinger Insitute for Mathematical Physics, and
by the Helmholtz program. A.Y.\ is grateful to M.~Bordag for his
kind hospitality at the University of Leipzig. 
\end{acknowledgments}

\end{document}